# Casimir repulsive-attractive transition between liquid-separated dielectric metamaterial and metal

Yingqi Ye, Qian Hu, Qian Zhao*, and Yonggang Meng

[1]State Key Laboratory of Tribology, Department of Mechanical Engineering,
Tsinghua University, Beijing, 100084, China
E-mail: zhaoqian@mail.tsinghua.edu.cn

**Abstract**

We study the repulsive-attractive transition, regarded as stable equilibrium, between gold and dielectric metamaterial based on Mie resonance immersed in various fluids due to the interplay of gravity, buoyancy and Casimir force among different geometries consisting of parallel plates and spheres levitated over substrates. A wider range of separation distance of stable equilibrium is obtained with Mie metamaterial than natural materials. We investigate the relationship between separation distance of stable equilibrium and constructive parameters of Mie metamaterial and geometric parameters of the system, and provide simple rules to tune the equilibrium position by modifying constructive parameters of Mie metamaterial. Particularly, the effect of permeability of Mie metamaterial on equilibrium separation is also considered. Our work is promising for potential applications in frictionless suspension in micro/nanofabrication technologies.

**Introduction**

The Casimir effect, which exists between neutral bodies at zero temperature due to vacuum fluctuations was first predicted by Casimir in 1948 [1]. Followed by his work, Lifshitz extended the theory from two semi-infinite perfectly electric conductor slabs to realistic dielectrics [2]. Due to the irreversible adhesion of neighboring elements in micro/nano-electromagnetic devices caused by Casimir attraction, how to alter attraction to repulsion becomes a hot point. One practical way to gain Casimir repulsion is to immerse two interacting objects in a fluid and satisfy $\varepsilon_1(i\xi) < \varepsilon_{\text{fluid}}(i\xi) < \varepsilon_2(i\xi)$ over a broad range of imaginary frequency $\xi$, where $\varepsilon_{\text{fluid}}(i\xi)$, $\varepsilon_1(i\xi)$ and $\varepsilon_2(i\xi)$ is dielectric permittivity of the fluid and two interacting bodies respectively [3]. Striking progress on theoretical calculations and experimental measurements on various liquids to obtain repulsive force has been made [4-7]. It is demonstrated in several works that a proper combination of two materials and a liquid can give rise to repulsive-attractive transition when electromagnetic characteristics of one of the interacting objects are close to the fluid over some range of frequency [8,9]. The unique behavior is desirable for potential applications in frictionless suspension and micro/nanofabrication technology. However, it occurs only in a few combinations of natural materials and the transition takes place in limited range of separation distance due to limited choices.

On the other hand, Casimir force can become repulsive if one of the two interacting objects is mainly electric while the other is mainly magnetic based on Boyer's theory [10]. In this way, the most challenging problem is to select or construct a material with sufficiently strong magnetic response and relatively low dielectric permittivity, which is almost impossible for natural magnetic materials. Numerous studies considering different artificial materials has been performed, including chiral metamaterial [11-13], hyperbolic metamaterial [14], electromagnetic metamaterial based on LC-circuit resonance [15,16] or Mie resonance [17] and other materials [18-24], providing a large amount of methods to minimize attraction but only a few theoretical predictions of repulsion.

Here, we propose a new approach to give rise to repulsive-attractive transition with a combination of dielectric metamaterial and metal immersed in a fluid, as the metamaterial with designable electric and magnetic properties is expected to be more suitable to meet the requirement for occurrence of transition. Here, metamaterial based on Mie resonance constructed by dielectric spherical particles dispersed into host media is chosen to display repulsive-attractive transitions for its relatively low dielectric permittivity and high magnetic permeability [25-28]. Furthermore, it can be treated as isotropic materials when spherical particles are identical and dispersed homogeneously [29]. The dielectric permittivity and magnetic permeability depend on constructive parameters, mainly including the filling factor of the particle in unit cell, size of particles and material of particles.

Schematics of plate-plate and sphere-plate geometries in vertical arrangement and horizontal arrangement are illustrated in Fig. 1. Levitation occurs with the balance of gravity, buoyancy and Casimir force in horizontal setup while only Casimir force is considered in vertical arrangement. In this paper, we analyze theoretically the deviation of transition position between Mie metamaterial slab and gold slab or sphere immersed in different liquids due to multiple parameters, including constructive parameters of Mie metamaterial and geometric parameters of the system in detail and find the rule to guide for designing Mie metamaterial to obtain a specific separation distance of equilibrium.

**Model and Theoretical Framework**

The Casimir force per unit area between two parallel plates immersed in a fluid separated by a distance $d$ at zero temperature can be expressed as

$$F_C = \frac{\hbar}{\pi} \int_0^\infty d\xi \iint \frac{d^2\mathbf{k}_\parallel}{(2\pi)^2} K_{\text{fluid}}$$

$$\times \sum_{p=\text{TE,TM}} \frac{r_1^p(i\xi,k) r_2^p(i\xi,k) e^{-2K_{\text{fluid}}d}}{1 - r_1^p(i\xi,k) r_2^p(i\xi,k) e^{-2K_{\text{fluid}}d}}$$

(1)

where the integral is carried out over imaginary frequency $\xi$ instead of real frequency $\omega$ to avoid singularities. $\hbar$ is the reduced Planck constant. $\mathbf{k}_\parallel$ is the wave vector parallel to surface, and $K_{\text{fluid}} = \sqrt{\varepsilon_{\text{fluid}}(i\xi)\xi^2/c^2 + k^2}$, where $c$ is the speed of light. $r_{1(2)}^p$ are diagonal elements of reflection matrices for two interacting objects, where off-diagonal elements are zero in our setup without entanglement of TE and TM mode. $r_{1(2)}^p$ can be written as

$$r_{1(2)}^{\text{TE}} = \frac{\mu_{1(2)}(i\xi)K_{\text{fluid}} - \sqrt{k_\parallel^2 + \varepsilon_{1(2)}(i\xi)\mu_{1(2)}(i\xi)\xi^2/c^2}}{\mu_{1(2)}(i\xi)K_{\text{fluid}} + \sqrt{k_\parallel^2 + \varepsilon_{1(2)}(i\xi)\mu_{1(2)}(i\xi)\xi^2/c^2}}$$

(2)

$$r_{1(2)}^{\text{TM}} = \frac{\varepsilon_{1(2)}(i\xi)K_{\text{fluid}} - \sqrt{k_\parallel^2 + \varepsilon_{1(2)}(i\xi)\mu_{1(2)}(i\xi)\xi^2/c^2}}{\varepsilon_{1(2)}(i\xi)K_{\text{fluid}} + \sqrt{k_\parallel^2 + \varepsilon_{1(2)}(i\xi)\mu_{1(2)}(i\xi)\xi^2/c^2}}$$

(3)

The Casimir force between a sphere of radius $R_s$ and a plate described by the proximity-force approximation (PFA), which is widely used in calculations of curved objects [30-32], can be written as $F_{\text{PFA}} = 2\pi R_s E$, where $E$ is the interaction energy per unit area in parallel-plates system composed of the same materials and at the same separation distance as in sphere-plate system. The effect of finite thickness of plate on Casimir force is also considered by replacing the reflection coefficients with generalized Fresnel coefficients [33,34].

We construct the Mie metamaterial by homogeneously dispersing dielectric spherical particles in air. Based on the Extended Maxwell Garnett (EMG) theory [29], the dielectric permittivity $\varepsilon_{\text{Mie}}$ and magnetic permeability $\mu_{\text{Mie}}$ of Mie metamaterial composed of particles of radius $R_p$ and filling factor $f$ at real frequency can be expressed as

$$\varepsilon_{\text{Mie}} = \frac{x^3 - 3ifT_1^E}{x^3 + \frac{3}{2}ifT_1^E} \tag{4}$$

$$\mu_{\text{Mie}} = \frac{x^3 - 3ifT_1^H}{x^3 + \frac{3}{2}ifT_1^H} \tag{5}$$

where $x = \omega R_p/c$, and $T_1^E$ and $T_1^H$ are the electric-dipole and magnetic-dipole coefficients of scattering matrix of a single particle, and can be given by the following formula:

$$T_1^E = \frac{j_1(x_p)[xj_1(x)]'\varepsilon_p(\omega) - j_1(x)[x_p j_1(x_p)]'}{h_1^+(x)[x_p j_1(x_p)]' - j_1(x_p)[xh_1^+(x)]'\varepsilon_p(\omega)} \tag{6}$$

$$T_1^H = \frac{j_1(x_p)[xj_1(x)]'\mu_p(\omega) - j_1(x)[x_p j_1(x_p)]'}{h_1^+(x)[x_p j_1(x_p)]' - j_1(x_p)[xh_1^+(x)]'\mu_p(\omega)} \tag{7}$$

where $x_p = \sqrt{\varepsilon_p(\omega)\mu_p(\omega)}R_p/c$, $\varepsilon_p(\omega)$ and $\mu_p(\omega)$ are permittivity and permeability of material of particles, respectively. $j_1$ is the spherical Bessel function in order one, $h_1^+$ is Hankel function in order one and $[zj_1(z)]' = d[zj_1(z)]/dz|_{z=x}$, etc. To apply permittivity and permeability of Mie metamaterial in Casimir force calculation formula, $\varepsilon_{\text{Mie}}(\omega)$ and $\mu_{\text{Mie}}(\omega)$ are converted to imaginary frequency domain according to Kramers-Kronig relation [2]:

$$\varepsilon(i\xi) = 1 + \frac{2}{\pi}\int_0^\infty \frac{\omega\varepsilon''(\omega)}{\xi^2 + \omega^2}d\omega \tag{8}$$

$$\mu(i\xi) = 1 + \frac{2}{\pi}\int_0^\infty \frac{\omega\mu''(\omega)}{\xi^2 + \omega^2}d\omega \tag{9}$$

where $\varepsilon''(\omega)$ and $\mu''(\omega)$ are the imaginary part of permittivity and permeability of the material calculated at real frequency respectively.

**Results and Discussion**

We investigate that at which combination of fluids and Mie metamaterial the transition of Casimir force from repulsion to attraction will occur first. Parallel-plates system in vertical arrangement is considered here, where gravity and buoyancy force are not taken into account. As mentioned above, repulsive Casimir force can be gained when $\varepsilon_1(i\xi) < \varepsilon_{\text{fluid}}(i\xi) < \varepsilon_2(i\xi)$ satisfied or an object mainly electric interacting with another object mainly magnetic. In our system, gold and Mie metamaterial are selected as the two interacting bodies. $\varepsilon_{\text{fluid}}(i\xi) < \varepsilon_{\text{gold}}(i\xi)$ is assumed to be satisfied over almost whole range of frequency since $\varepsilon_{\text{gold}}(i\xi)$ is rather large, remaining $\varepsilon_{\text{Mie}}(i\xi) < \varepsilon_{\text{fluid}}(i\xi)$ to be examined. Gold and all of the liquids selected here are non-magnetic with $\mu = 1$, and the permeability of Mie metamaterial depends on its constructive parameters. The dielectric permittivity of several kinds of fluids and Mie metamaterial composed of silicon spherical particles with radius $R_p = 100\text{nm}$ and filling factor $f = 0.5$ is shown in Fig. 2(a), and the permeability of Mie metamaterial at imaginary frequency is plotted in the inset. The requirement $\varepsilon_{\text{Mie}}(i\xi) < \varepsilon_{\text{fluid}}(i\xi)$ is satisfied or violated over different range of frequency, giving rise to one crossing from curve of Mie metamaterial with bromobenzene and two crossings with ethanol and pure water. As pointed out in [9], type I crossing, which converts $\varepsilon_{\text{Mie}}(i\xi) > \varepsilon_{\text{fluid}}(i\xi)$ to $\varepsilon_{\text{Mie}}(i\xi) < \varepsilon_{\text{fluid}}(i\xi)$ from lower frequency to higher frequency, accounts for the repulsive-attractive transition of Casimir force, since the lower frequency range dominates the contribution to Casimir force at further distance while the higher frequency range dominates at closer distance. The transition from attraction to repulsion arises from type II crossing, which converts $\varepsilon_{\text{Mie}}(i\xi) < \varepsilon_{\text{fluid}}(i\xi)$ to $\varepsilon_{\text{Mie}}(i\xi) > \varepsilon_{\text{fluid}}(i\xi)$ from lower frequency to higher frequency. The repulsive-attraction transition is regarded as stable equilibrium, since two slabs will be pushed together by attractive force when separating and be pulled apart by repulsive force when approaching, while the transition from attraction to repulsion is unstable. For the other way to achieve repulsion, gold is regarded as the electric one and Mie metamaterial as the magnetic one. Repulsive Casimir force is strengthened and stable equilibrium occur at larger separation with increasing permeability of Mie metamaterial, which will be discussed more detailedly in the following context.

Figure 2(b) shows the Casimir force between two semi-infinite plates of gold and Mie metamaterial separated by these fluids, normalized by the Casimir force between two perfectly electric conductors in vacuum, which is purely attractive. An interesting behavior observed here is that, two, one and zero transitions take place in ethanol, bromobenzene and pure water respectively, corresponding to two, one and two crossings in Fig. 2(a). The stable equilibrium at separation $d \approx 132\text{nm}$ and unstable equilibrium at separation $d \approx 1.24\text{μm}$ are found in ethanol as a consequence of one type I crossing and one type II crossing in curve of $\varepsilon_{\text{Mie}}(i\xi)$ and $\varepsilon_{\text{ethanol}}(i\xi)$. The repulsive-attractive transition at separation $d \approx 127\text{nm}$ in bromobenzene comes from contribution by type I crossing in curve of $\varepsilon_{\text{Mie}}(i\xi)$ and $\varepsilon_{\text{bromobenzene}}(i\xi)$. Casimir force in water is purely repulsive, due to the fact that, the violation of $\varepsilon_{\text{Mie}}(i\xi) < \varepsilon_{\text{fluid}}(i\xi) < \varepsilon_{\text{gold}}(i\xi)$ occurs over not sufficiently wide range of frequency, overwhelmed by the satisfaction of requirement for achieving repulsion. Therefore, the rough rule to predict transition by crossings of curve of electromagnetic parameters alone may lead to incorrect results, but still can guide us. Only stable equilibrium is discussed in the following parts, which is more desirable for potential applications.

To investigate the relationship between separation distance of stable equilibrium and constructive parameters of Mie metamaterial, we study the effect of constructive parameters variations on permittivity and permeability of Mie metamaterial first. The permittivity and permeability for different filling factor $f$, different particle radius $R_\text{p}$ and particles made of different materials are numerically calculated and illustrated in Fig. 3. The permittivity increases with increasing filling factor in Fig. 3(a), similar increase in permeability as filling factor increases is observed in Fig. 3(b). The increase of permittivity occurs much more rapidly than permeability, contributing to more Casimir attraction. Comparing to the permittivity of ethanol, repulsive-attractive transition is expected to take place among the range of filling factor from 0.4 to 0.6. For different particle radius, oppositely, permittivity decreases as the size of particles increases in Fig. 3(c). The effect of size of particles on permeability is much more complicated as shown in Fig. 3(d). Permeability is strengthened in lower part of frequency but weakened in higher part with increasing size of particle. A feature worth noting is that, both of the permittivity and permeability varies more dramatically under changing of filling factor compared with the size of particles, leading to greater deviation of equilibrium position by modifying filling factor rather than the size of particles. In Fig. 3(e) and (f), we numerically calculate permittivity and permeability of Mie metamaterial composed of three different dielectric particles:

teflon, silica and silicon spheres. The permittivity is the highest and permeability is the lowest when made of teflon particles, and vice versa for silicon particles. Therefore, silicon is expected as a better candidate for achieving repulsive-attractive transition.

The relative Casimir force $F/F_{\text{PEC}}$ between ethanol-separated gold slab and Mie metamaterial slab for different filling factor $f$, different particle radius $R_{\text{p}}$ and particles of different materials is numerically calculated, as plotted in Fig.4, where $F_{\text{PEC}}$ is the force between two perfectly electric conductors in vacuum. We investigate the effect of permittivity of Mie metamaterial on equilibrium first. As mentioned above, stable equilibrium may arise from a type I crossing. The rough rule can be extended as, a type I crossing at lower frequency means a stable equilibrium at further distance. Because more repulsion is contributed by permittivity at high frequency as the requirement $\varepsilon_{\text{Mie}}(i\xi) < \varepsilon_{\text{fluid}}(i\xi) < \varepsilon_{\text{gold}}(i\xi)$ met under the shift of the crossing to lower frequency. As expected, transition from repulsion to attraction occurs when $f \approx 0.5$ as shown in Fig. 4(a). The equilibrium position deviates to further separation distance with decreasing filling factor, corresponding to the variation of type I crossing of curve of $\varepsilon_{\text{Mie}}(i\xi)$ and $\varepsilon_{\text{ethanol}}(i\xi)$ to lower frequency under smaller filling factor. Similar change of equilibrium position to further separation distance take place in Fig. 4(b), as the size of particles increases, mainly caused by the shift of the type I crossing to the left with increasing size of particles. Specially, the deviation of stable equilibrium position is dramatically sensitive to the change of filling factor, which should be treated carefully in applications of Mie metamaterial in micro/nano-electromagnetic devices. Another interesting phenomenon observed in Fig. 4(c) and (f) is that, although the crossings of curve of ethanol with silicon, teflon and silica particle occur nearby, the equilibrium position of silicon particle differs from teflon and silica, mainly due to the contrast between permittivity of silicon particle with teflon and silica in the range of lower frequency. Since an accurate result is carried out over whole range of frequency, the rule to predict the position of transition by the position of crossing of curve of permittivity alone fails in some situations, but is still helpful. We obtain a wider range of equilibrium separation distance from 50nm to 300nm for all the parameters evaluated here, compared by 50-200nm obtained by natural materials [9], and equilibrium separation of a few micros can be reached with larger filling factor. A larger separation is more convenient for experimental observation. Furthermore, a specific separation distance can be obtained by selecting appropriate constructive parameters of Mie metamaterial. A rough conclusion drawn here is that, the stable equilibrium position can be tuned to further separation distance by decreasing filling factor, increasing size of particles and selecting silicon instead of teflon or silica. For all parameters investigated, the equilibrium position is the most sensitive to filling factor.

To extract the effect of permeability on stable equilibrium position, a hypothetical material of the same permittivity with Mie metamaterial and $\mu = 1$ is proposed. The stable equilibrium position of ethanol-separated gold and Mie metamaterial and the hypothetical material for different constructive parameters is compared, as plotted in Fig. 5. For gold and the hypothetical material, the stable equilibrium occurs at closer distance. As mentioned above, Casimir repulsion is achieved between a plate mainly electric and another plate mainly magnetic according to Boyer's theory. In our setup, gold is the electric one, and more repulsion is contributed by Mie metamaterial than the hypothesis material due to the larger permeability of Mie metamaterial. Consequently, the repulsive-attractive transition occur at further distance in gold and Mie metamaterial than the hypothetical material for all constructive parameters. As gold and the fluids selected here are all non-magnetic, the equilibrium position changes to further distance with increasing permeability of Mie metamaterial.

We now consider the stable levitation in the system at which a thin gold film suspended over Mie metamaterial substrate immersed in a fluid, by the balance of gravity, buoyancy and Casimir force. The gravity and buoyancy force scale as the thickness of gold film, as the surface area of gold film is fixed. The density of gold, ethanol and bromobenzene: $\rho_{\text{gold}}$, $\rho_{\text{ethanol}}$ and $\rho_{\text{bromobenzene}}$, takes the value: 19.26g/cm$^3$, 0.789g/cm$^3$ and 1.5g/cm$^3$. The separation distance of stable equilibrium of gold film suspended over Mie metamaterial substrate separated by ethanol for different filling factor $f$, different particle radius $R_{\text{p}}$ and different materials is numerically calculated, when film thickness is fixed to 200nm, as plotted in Fig. 6(a), (b) and (c). The law of equilibrium separation distance changing with all these parameters is the same as in vertical arrangement, but all of the equilibrium separations move to closer distance compared with the system in vertical arrangement. Because repulsive Casimir force decays rapidly with increasing distance and gravity force becomes the dominant one and turn the resultant force from repulsion to attraction at closer distance. A feature worth noting is that, the equilibrium separation distance changes less sensitive with all the parameters investigated in horizontal arrangement, which can be used to modify the equilibrium separation distance more precisely. In Fig. 6(d), the total force of gold film suspended over Mie metamaterial substrate immersed in different fluids is plotted. Repulsive-attractive transition take place in water-separated system at separation distance $d \approx 135$nm, while the force is purely repulsive and no transition occurs in vertical arrangement, as a result of gravity too.

The effect of variation of constructive parameters of Mie metamaterial on equilibrium position is similar in sphere-plate system in vertical arrangement and horizontal arrangement. Then we turn to the effect of geometric parameters of parallel-plates and sphere-plate system on equilibrium position, including film thickness and sphere radius. For parallel-plates system, the change of equilibrium position as a function of film thickness of gold in system of vertical arrangement and horizontal arrangement is shown in Fig. 7(a). In vertical arrangement, the equilibrium is reached at further distance with increasing thickness of gold film and asymptotic to the value of semi-infinite gold plate at large thickness. Because repulsive Casimir force is weakened by decreasing gold film thickness. In horizontal arrangement, the equilibrium separation increases to a peak and then go down as film thickness increases. The upward trend at closer distance is contributed by repulsive Casimir force, and downward trend by gravity force. For sphere-plate system in vertical arrangement, a unique phenomenon observed is that, all transitions occur at the same distance with different sphere radius, as shown in Fig. 7(b), due to the fact that, Casimir force increases or decreases proportional to the change of sphere radius and the equilibrium position stay invariant by PFA method. It seems possible to assembly nanoparticles in a regular sequence by the same equilibrium position, however, deviations of PFA method from accurate results can not be negligible when the curvature of sphere is close to separation distance [35,36]. In the system of a gold sphere levitated by Mie metamaterial substrate, the separation distance of stable suspension decreases as sphere radius increases, because the gravity grows much more rapidly than repulsive Casimir force with increasing sphere radius.

**Conclusion**

We have performed a theoretical analysis on the equilibrium phenomenon between liquid-separated gold and Mie metamaterial among various geometries. A wider range of equilibrium separation is obtained with Mie metamaterial than natural materials.

Stable levitation of gold film and gold sphere over Mie metamaterial substrates immersed in different fluids is also obtained. Roughly, the levitation position can be tuned to larger separation by decreasing the filling factor or increasing the size of particle. A larger separation is also reached with a larger permeability of Mie metamaterial. Among all the parameters, the equilibrium position is the most sensitive to the filling factor. The effect of geometric parameters of the system in vertical arrangement and horizontal arrangement are compared. The gravity force becomes the dominant one with increasing size of bodies, but the interplay between gravity and Casimir force of tiny bodies is complicated. Our results serve as a guide for applications of Mie metamaterial in frictionless suspension in micro/nanofabrication technologies.

**Acknowledgements**

This work is supported by the National Natural Science Foundation of China (Grant Nos. 51575297, 51635009), the Science and Technology Plan of Shenzhen City (JCYJ20160301154309393), and the Chinese State Key Laboratory of Tribology.

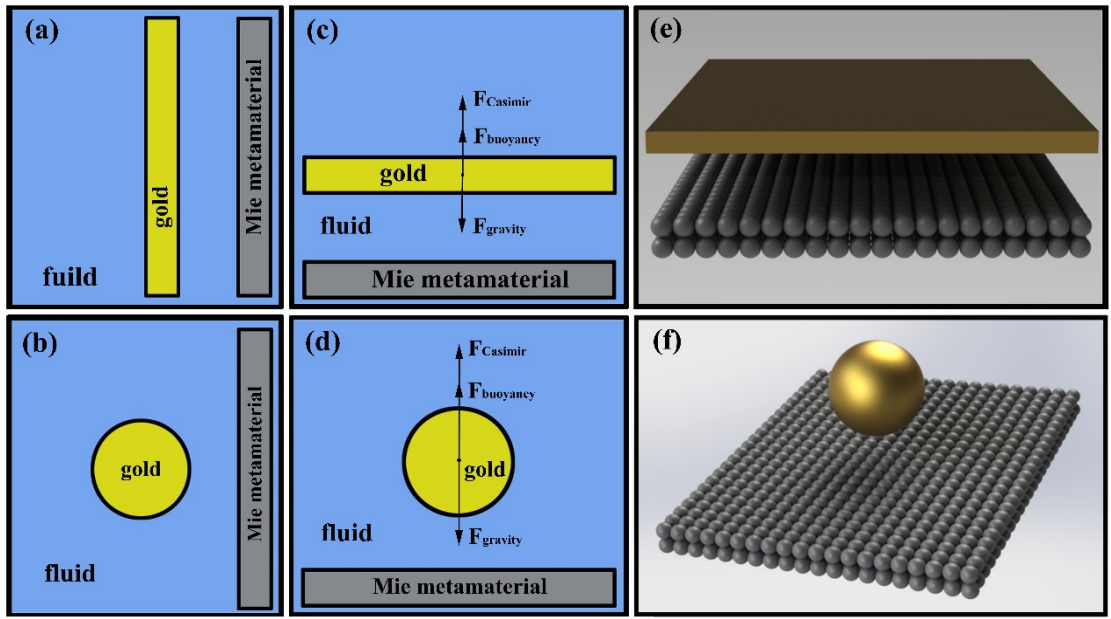

Fig. 1. Schemes for liquid-separated gold and Mie metamaterial among different geometries: (a) parallel-plates system in vertical arrangement, (b) sphere-plate system in vertical arrangement, (c) parallel-plates system in horizontal arrangement and (d) sphere-plate system in horizontal arrangement. Schemes for (e) a gold film or (f) a gold sphere suspended over Mie metamaterial substrate.

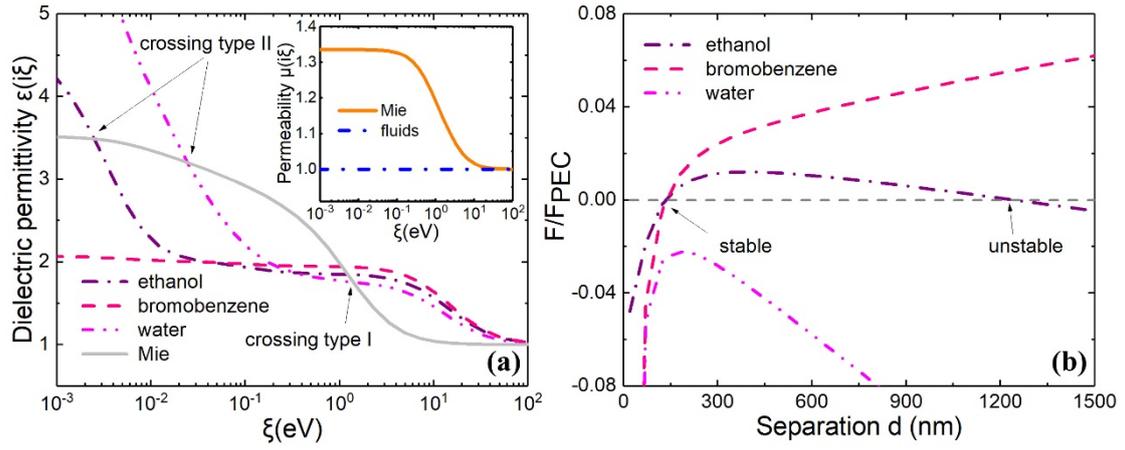

Fig. 2. (a) Dielectric permittivity of ethanol, bromobenzene, pure water and Mie metamaterial made of silicon particles ($R_\mathrm{p} = 100\mathrm{nm}$, $f = 0.5$) at imaginary frequency. (b) Casimir force between gold and Mie metamaterial semi-infinite plate immersed in ethanol, bromobenzene and pure water, normalized by $F_\mathrm{PEC}$, the Casimir force between perfectly electric conductors in vacuum.

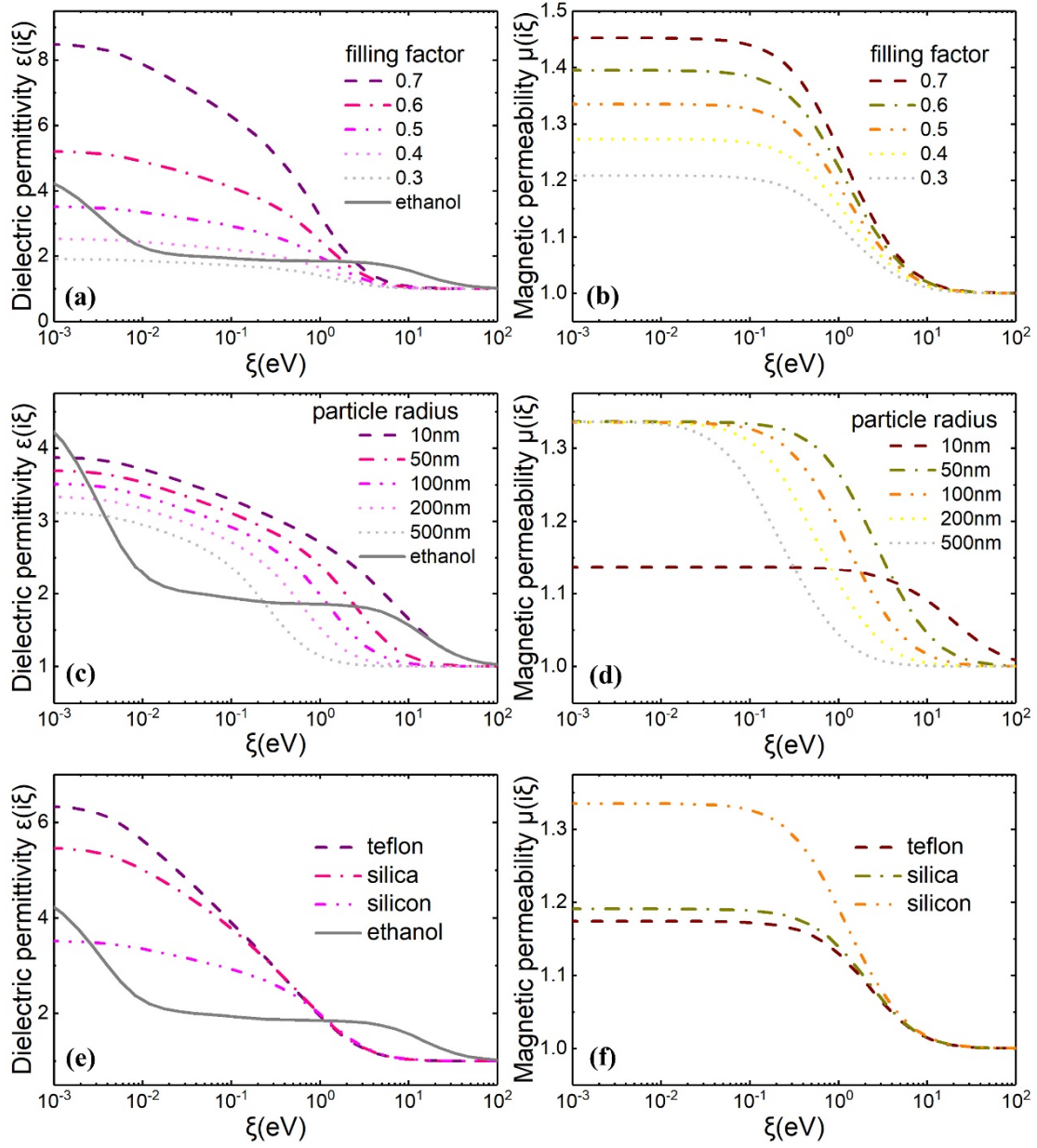

Fig. 3. (a) Permittivity and (b) permeability of Mie metamaterial for different filling factor $f$ when made of silicon particles, $R_p = 100$nm. (c) Permittivity and (d) permeability of Mie metamaterial for different particle radius $R_p$ when made of silicon particles, $f = 0.5$. (e) Permittivity and (f) permeability of Mie metamaterial for particles of different materials, $R_p = 100$nm, $f = 0.5$.

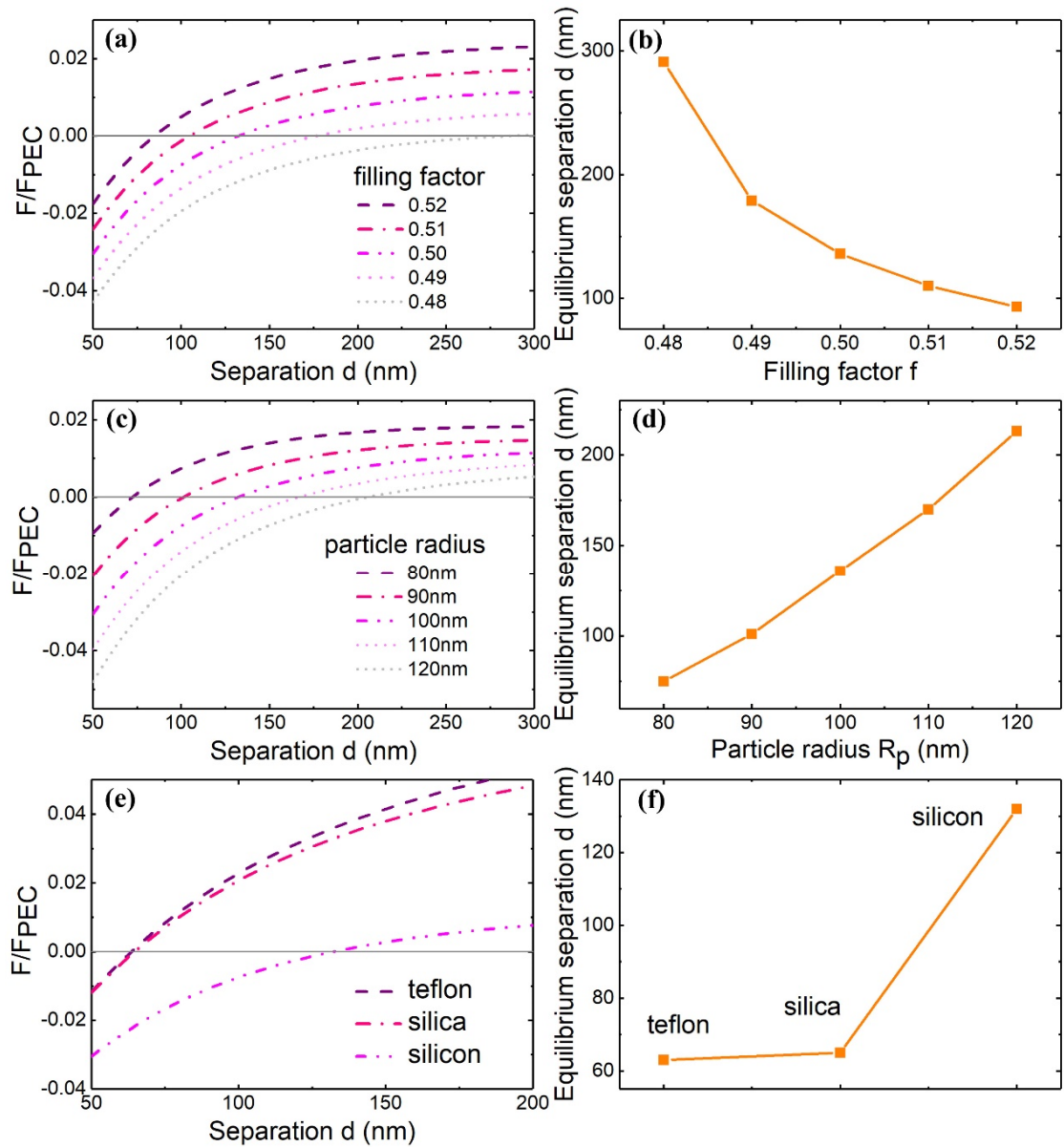

Fig. 4 Casimir force between gold and Mie metamaterial semi-infinite plate immersed in ethanol for (a) different filling factor $f$, when made of silicon particles, $R_p = 100$nm (b) different particle radius $R_p$, when made of silicon particles, $f = 0.5$ and (c) different materials when $R_p = 100$nm and $f = 0.5$, normalized by $F_{PEC}$, the Casimir force between perfectly electric conductors in vacuum. (d)-(f) Equilibrium separation distance in (a)-(c), respectively.

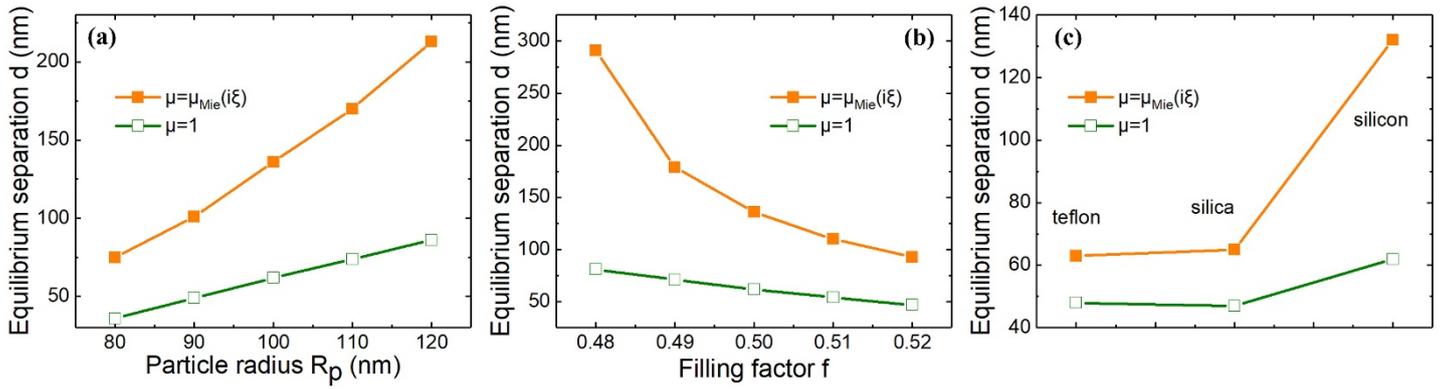

Fig.5 Equilibrium separation distance of gold and Mie metamaterial(hypothetical material) semi-infinite plate immersed in ethanol for (a) different filling factor $f$, (b) different particle radius $R_p$ and (c) different materials, all the other parameters are the same as in Fig. 4 (a)-(c).

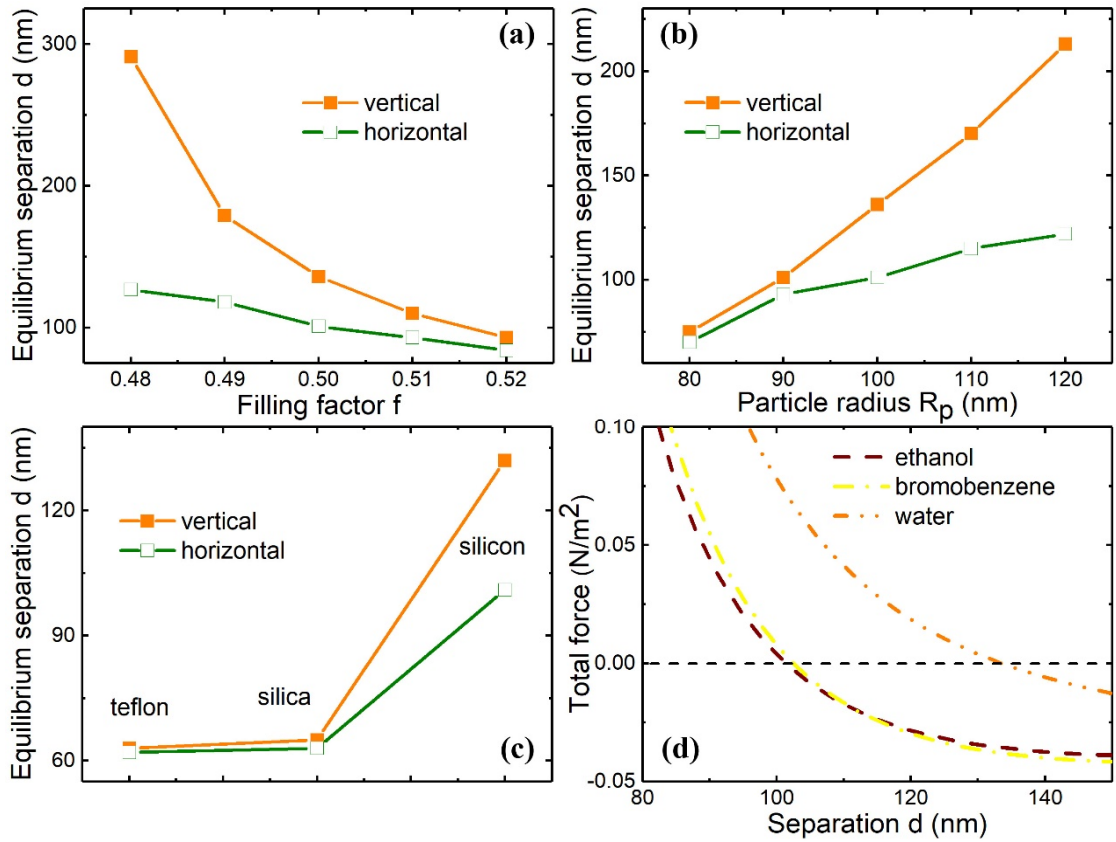

Fig.6 Equilibrium separation distance of gold and Mie metamaterial semi-infinite plate immersed in ethanol of the system in vertical(horizontal) arrangement for (a) different filling factor $f$, (b) different particle radius $R_p$ and (c) different materials, all the other parameters are the same as in Fig. 4(a)-(c). (d) Total force of gold and Mie metamaterial semi-infinite plate immersed in different fluids of the system in horizontal arrangement, all the other parameters are the same as in Fig. 2(a).

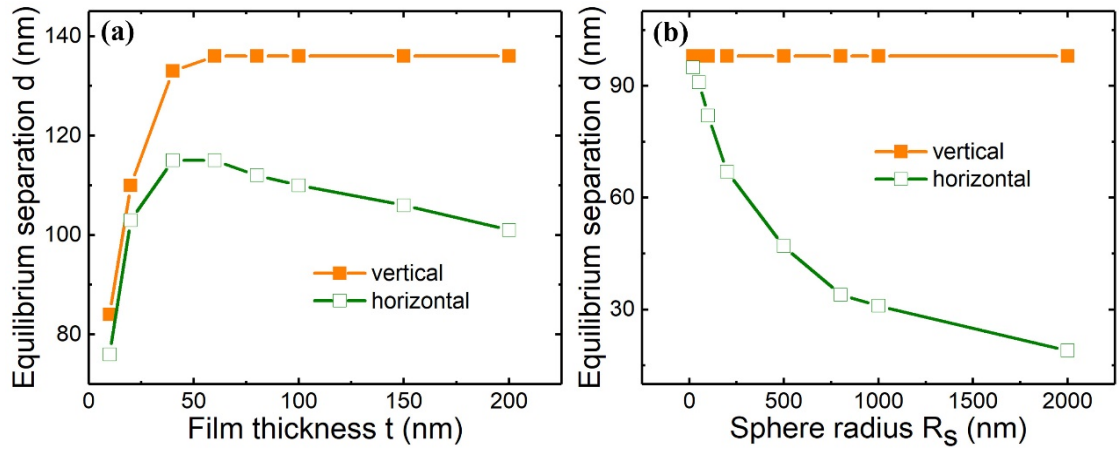

Fig.7 Equilibrium separation distance of gold and Mie metamaterial immersed in ethanol of (a) the parallel-plate system in vertical(horizontal) arrangement for different film thickness $t$, (b) the sphere-plate system in vertical(horizontal) arrangement for different sphere radius $R_s$, all the other parameters are the same as in Fig. 2(a).